\documentclass[aps,prl,reprint, superscriptaddress, bibliography]{revtex4-1}
\usepackage[section]{placeins}

\usepackage{lineno}
\usepackage{graphicx}  
\usepackage{dcolumn}   
\usepackage{bm}        
\usepackage{amssymb}   
\usepackage{amsmath,stmaryrd}
\usepackage{blkarray, multirow, graphicx, diagbox, color, colortbl}
\usepackage[dvipsnames]{xcolor}
\usepackage{bbm, bbold}
\usepackage{glossaries}
\usepackage{hyphenat}
\usepackage{ifthen}
\usepackage[colorlinks, linkcolor = blue, citecolor = blue, filecolor = black, urlcolor = blue]{hyperref}
\usepackage{xkeyval}
\usepackage{moreverb}
\usepackage{rotating}
\usepackage{wrapfig}
\usepackage{slashbox}
\usepackage{xspace}
\usepackage{nicefrac}
\usepackage[]{units}
\usepackage{physics}
\usepackage{booktabs}
\usepackage{braket}
\usepackage[inline]{enumitem}
\usepackage{tabto}
\usepackage{listings}
\usepackage{xstring}
\usepackage{pgfplots}
\pgfplotsset{compat=1.18}
\def\ReplaceStr#1{%
	\IfSubStr{#1}{p}{%
		\StrSubstitute{#1}{p}{.}}{#1}}

\usepackage[caption=false]{subfig}
\usepackage[capitalise]{cleveref} %
\usepackage{chemformula}
\usepackage{algorithm}

\newcommand{\nodagger}[0]{{\vphantom{\dagger}}}

\newcommand{\coherence}[2]{\mathrm{CSO}_{#2}[#1]}

\newacronym{OBC}{OBC}{open boundary condition}
\newacronym{PBC}{PBC}{periodic boundary condition}
\newacronym{TFIM}{TFIM}{transverse-field Ising model}
\newacronym{RDM}{RDM}{reduced density matrix}
\newacronym{NS}{NS}{Neveu-Schwarz}
\newacronym{R}{R}{Ramond}
\newacronym{JW}{JW}{Jordan\hyp Wigner}
\newacronym{CSO}{CSO}{coherent\hyp subspace overlap}
\newacronym{RMS}{RMS}{repeated\hyp measurement statistics}
\graphicspath{{figures/}}
\begin{document}
\author{Leonard Werner Pingen} 
\affiliation{Department of Physics, Arnold Sommerfeld Center for Theoretical Physics (ASC), Munich Center for Quantum Science and Technology (MCQST), Ludwig-Maximilians-Universit\"{a}t M\"{u}nchen, 80333 M\"{u}nchen, Germany}

\author{Mattia Moroder}\thanks{These authors contributed equally} 
\affiliation{School of Physics, Trinity College Dublin, Dublin 2, Ireland}

\author{Sebastian Paeckel} \thanks{These authors contributed equally} 
\affiliation{Department of Physics, Arnold Sommerfeld Center for Theoretical Physics (ASC), Munich Center for Quantum Science and Technology (MCQST), Ludwig-Maximilians-Universit\"{a}t M\"{u}nchen, 80333 M\"{u}nchen, Germany}

\def\thetitle{Probing the Physical Reality of Projective Measurements}
\title{\thetitle}
\begin{abstract}
We propose a protocol to test whether the postulate of a measurement acting as an instantaneous projection onto an eigenstate of the measurement apparatus is compatible with physical reality.
This approach is solely based on repeated measurements of local quantities with frequencies that are within reach of analog quantum simulation platforms, for instance Rydberg atom arrays or ultracold gases in optical lattices.
Crucially, we also develop a continuous description of a quantum measurement finding that its~\gls{RMS} drastically differ from the projective case.
This description is based on very general assumptions about quantum systems, most importantly maintaining continuous dynamics of the coherent part of the state.
Our findings imply that the significantly different measurement statistics in the collapse\hyp free description should be qualitatively replicated by any modification of standard quantum theory that is lacking explicit wave\hyp function collapses.
\end{abstract}
\maketitle
\begin{figure*}[htb]
    \centering
    \includegraphics[width=\linewidth]{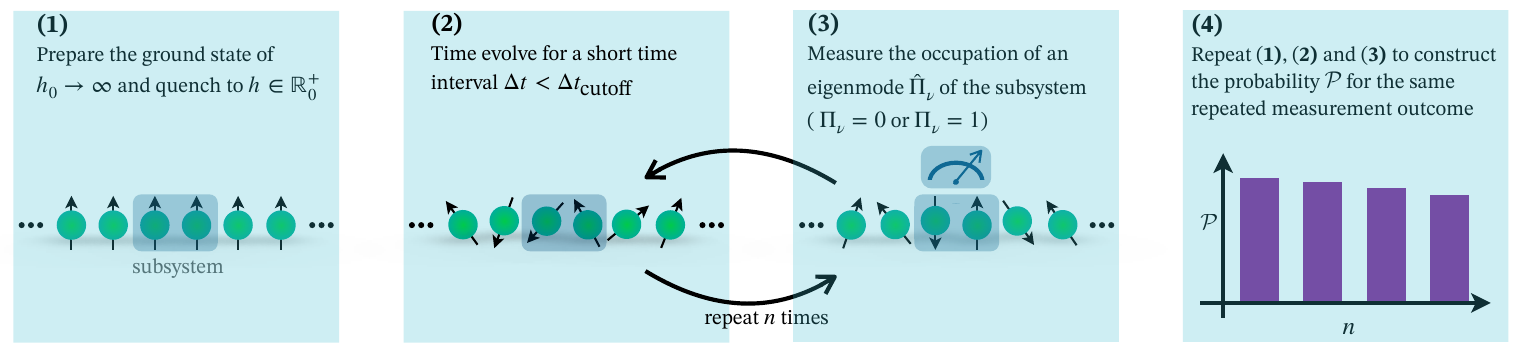}
    \caption{
        The repeated\hyp measurement protocol.
        We initially prepare a chain of aligned spin-$1/2$ particles (1) and then evolve it for a short time with the \gls{TFIM} Hamiltonian \cref{eq:TFIM:Hamiltonian} with a finite field strenght $h$ (2).
        For a subsystem composed of $N$ spins, we measure $\hat{\Pi}_\nu$, which is a conserved quantity of the isolated subsystem (3).
        We alternate short time evolutions and subsystem measurements for $n$ timesteps.
        Finally, we repeat steps (1), (2) and (3) to compute the probability of confirming the initial measurement outcome, i.e. $\Pi_\nu=0$ or $\Pi_\nu=1$, $n$ consecutive times (4). Projective measurements and continuous measurements yield different \gls{RMS} (see \cref{fig:measurement_statistics}).
    }
    \label{fig:first:figure}
\end{figure*}
\begin{figure}
    \centering
    \resizebox{\linewidth}{!}{
        \begin{tikzpicture}[
            every axis/.style={
                    ybar,
                    ymin=0,ymax=1,
                    axis y line=left,
                    ytick style={/pgfplots/major tick length=3pt},
                    ytick style={xshift=-1.5pt},
                    xtick=data,
                    x tick label style={anchor=north},
                    xtick style={draw=none},
                    symbolic x coords={1,2,3,4,5,6,7,8,9,10},
                    axis x line=bottom,
                    axis line style={-},
                    x=34pt,
                    bar width=8pt,
                    enlarge x limits=0.06
                },
            ]
            \tikzset{every node/.style={font=\fontsize{13pt}{13pt}\selectfont}}
            
            \begin{axis}[
                bar shift=-9pt,
                hide axis,
                legend cell align={center},
                legend style={at={(0.5,1.02)}, anchor=south, draw=none, legend columns = -1, column sep=2pt},
                legend image post style={xscale=0.6},
            ]
        
                \addlegendimage{ybar, fill=Plum, draw opacity=0, yshift=-2pt}
                \addlegendentry{$\mathcal{P}_\mathrm{proj}^\mathrm{min}(w_n=1) ~~~~$}
                \addlegendimage{ybar, fill=Salmon, draw opacity=0, yshift=-2pt}
                \addlegendentry{$\tilde{\mathcal{P}}_\mathrm{proj}^\mathrm{min}(w_n=1) ~~~~$}
                \addlegendimage{ybar, fill=Turquoise, draw opacity=0, yshift=-2pt}
                \addlegendentry{$\mathcal{P}_\mathrm{dec}^\mathrm{max}(w_n=1)$}
                        
                \addplot+[fill=Plum, draw=none] coordinates
                {(1,0.99) (2,0.9801) (3,0.970299) (4,0.96059601) (5,0.9509900498999999) (6,0.941480149401) (7,0.9320653479069899) (8,0.9227446944279201) (9,0.9135172474836408) (10,0.9043820750088044)};
            \end{axis}
            
            \begin{axis}[
                bar shift=-0pt,
                hide axis
            ]
            \addplot+[fill=Salmon, draw=none] coordinates
            {(1,0.52901516) (2,0.27985704) (3,0.14804862) (4,0.07831996) (5,0.04143245) (6,0.02191839) (7,0.01159516) (8,0.00613402) (9,0.00324499) (10,0.00171665)};
            \end{axis}
            
            \begin{axis}[
                bar shift=9pt,
                xlabel={Confirming measurement number $n$ at time $t_n = n t_\mathrm{bound}$}
            ]
            \addplot+[fill=Turquoise, draw=none] coordinates
            {(1,0.26500528) (2,0.07045371) (3,0.01883036) (4,0.00506992) (5,0.00137776) (6,0.00037860) (7,0.00010538) (8,0.00002976) (9,0.00000854) (10,0.00000249)};
            \end{axis}
        
            \begin{scope}[xshift=234pt, yshift=80pt] 
                \node[fill=white, draw=black, thin, rounded corners=.5pt] at (0,0) [minimum width=7.2cm, minimum height=4cm] {}; 
            \end{scope}
        
            \begin{scope}[xshift=330pt, yshift=132pt] 
                \begin{axis}[
                    ybar,
                    bar shift=-4pt,
                    width=4cm, height=4.8cm, 
                    at={(0,0)}, anchor=north east, 
                    xmode=normal,
                    xmin=3, xmax=8,
                    ymin=0.00001,
                    ymax=0.15,
                    xtick={3,4,5,6,7,8},
                    ytick={0.00001, 0.0001, 0.001, 0.01, 0.1},
                    yticklabels={\(\), $10^{-4}$, $10^{-3}$, $10^{-2}$, $10^{-1}$},
                    x tick label style={yshift=2pt},
                    enlarge x limits=0.1,
                    x=28pt,
                    ymode=log,
                    log origin=infty,
                ]
                    
                    \addplot+[fill=Salmon, draw=none] coordinates
                    {(3,0.14804862) (4,0.07831996) (5,0.04143245) (6,0.02191839) (7,0.01159516) (8,0.00613402)};
                \end{axis}
        
                \begin{axis}[
                    ybar,
                    bar shift=0pt,
                    hide axis,
                    width=4cm, height=4.8cm,
                    at={(0,0)}, anchor=north east,
                    xmode=normal,
                    xmin=3, xmax=8,
                    ymin=0.00001,
                    ymax=0.15,
                    x=28pt,
                    ymode=log,
                    log origin=infty,
                ]
                    \addplot+[fill=Turquoise, draw=none] coordinates
                    {(3,0.01883036) (4,0.00506992) (5,0.00137776) (6,0.00037860) (7,0.00010538) (8,0.00002976)};
                \end{axis}
            \end{scope}
        \end{tikzpicture}
    }
    \caption{
    Lower ($\mathcal{P}_\mathrm{proj}^\mathrm{min}, \tilde{\mathcal{P}}_\mathrm{proj}^\mathrm{min}$) and upper ($\mathcal{P}_\mathrm{dec}^\mathrm{max}$) bounds for the relative incidence $w_n$ to confirm an initial measurement outcome $\Pi$ for $n$ consecutive times under the assumption of projective (purple, salmon colored bars) or decoherence\hyp based measurements (turquoise bars), following a quench inside the disordered phase of the~\gls{TFIM}.
    We set $P_\mathrm{min} = 0.99$ and the subsystem to $N=8$ sites.
    The resulting inter\hyp measurement timescale is $t_\mathrm{bound} \approx 0.0152/J$, which implies $t_\mathrm{bound} \in \mathcal{O} (1 ~ m\mathrm{s})$ for practical setups~\cite{Meinert2013}.
    $\mathcal{P}_\mathrm{proj}^\mathrm{min}, \mathcal{P}_\mathrm{dec}^\mathrm{max}$ were derived for a direct measurement of the quasi\hyp conserved quantities of the subsystem, while $\tilde{\mathcal{P}}_\mathrm{proj}^\mathrm{min}$ was obtained using measurements of the spin\hyp $z$ projection, only.
    The inset shows $\mathcal{P}_\mathrm{dec}^\mathrm{max}$ and $\tilde{\mathcal{P}}_\mathrm{proj}^\mathrm{min}$ in logarithmic scale.
    }
    \label{fig:measurement_statistics}
\end{figure}
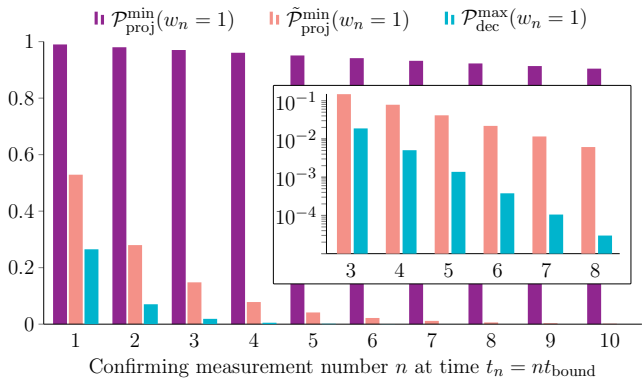
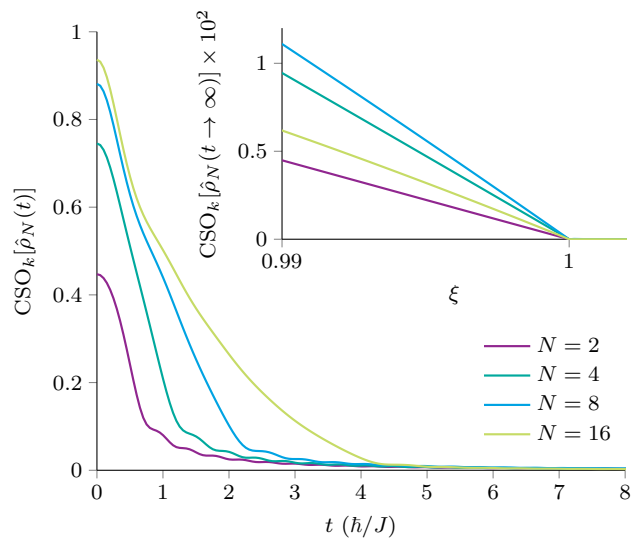
\begin{figure}[t]
    \centering
    \resizebox{\linewidth}{!}{
        \begin{tikzpicture}
            \begin{axis}[
                ymin=0, ymax=1,
                xmin=0, xmax=8,
                axis y line=left,
                xtick style={/pgfplots/major tick length=3pt},
                xtick style={yshift=-1.5pt},
                ytick style={/pgfplots/major tick length=3pt},
                ytick style={xshift=-1.5pt},
                xtick={0, 1, 2, 3, 4, 5, 6, 7, 8},
                xlabel={$t ~ (\hbar / J)$},
                ylabel={$\coherence{\hat\rho_N (t)}{k}$},
                legend pos=north east,
                legend cell align=left,
                legend style={
                    draw=none,
                    at={(1,0.035)},
                    anchor=south east
                },
                axis x line=bottom,
                axis line style={-}
            ]
            \tikzset{every node/.style={font=\fontsize{8pt}{8pt}\selectfont}}
        
            \addplot [
                solid,
                mark=none,
                thick,
                draw=Plum
            ] table [
                x index=0,
                y index=1,
                col sep=space
            ] {data/coherence_n_2.dat};
            \addlegendentry{$N = 2$}
        
            \addplot [
                solid,
                mark=none,
                thick,
                draw=Emerald
            ] table [
                x index=0,
                y index=1,
                col sep=space
            ] {data/coherence_n_4.dat};
            \addlegendentry{$N = 4$}
        
            \addplot [
                solid,
                mark=none,
                thick,
                draw=Cerulean
            ] table [
                x index=0,
                y index=1,
                col sep=space
            ] {data/coherence_n_8.dat};
            \addlegendentry{$N = 8$}
        
            \addplot [
                solid,
                mark=none,
                thick,
                draw=SpringGreen
            ] table [
                x index=0,
                y index=1,
                col sep=space
            ] {data/coherence_n_16.dat};
            \addlegendentry{$N = 16$}
            
            \end{axis}
            
            \begin{scope}[shift={(2.4, 3)}]
                \begin{axis}[
                    width=.7\linewidth, height=.5\linewidth,
                    ymin=0, ymax=1.2,
                    xmin=0.99, xmax=1.002,
                    axis y line=left,
                    xtick style={/pgfplots/major tick length=3pt},
                    xtick style={yshift=-1.5pt},
                    tick label style={font=\footnotesize},
                    ytick style={/pgfplots/major tick length=3pt},
                    ytick style={xshift=-1.5pt},
                    xtick={0.99, 1},
                    xlabel={$\xi$},
                    xlabel style={font=\footnotesize},
                    ylabel={$\coherence{\hat\rho_N (t \to \infty)}{k} \times 10^{2}$},
                    ylabel style={font=\footnotesize},
                    axis x line=bottom,
                    axis line style={-}
                ]
                    \addplot[
                        solid,
                        mark=none,
                        thick,
                        draw=Plum
                    ] table [
                        x index=0,
                        y index=1,
                        y expr=\thisrowno{1}*100,
                        col sep=space
                    ] {data/limiting_coherence_n_2.dat};
        
                    \addplot[
                        solid,
                        mark=none,
                        thick,
                        draw=Emerald
                    ] table [
                        x index=0,
                        y index=1,
                        y expr=\thisrowno{1}*100,
                        col sep=space
                    ] {data/limiting_coherence_n_4.dat};
        
                    \addplot[
                        solid,
                        mark=none,
                        thick,
                        draw=Cerulean
                    ] table [
                        x index=0,
                        y index=1,
                        y expr=\thisrowno{1}*100,
                        col sep=space
                    ] {data/limiting_coherence_n_8.dat};
        
                    \addplot[
                        solid,
                        mark=none,
                        thick,
                        draw=SpringGreen
                    ] table [
                        x index=0,
                        y index=1,
                        y expr=\thisrowno{1}*100,
                        col sep=space
                    ] {data/limiting_coherence_n_16.dat};
                \end{axis}
                
            \end{scope}
        \end{tikzpicture}
    }
    \caption{
        \Acrfull{CSO} for a quench to $\xi = 1$ and different subsystem sizes $N$.
        The measured quasi\hyp conserved quantities for the individual values of $N$ are chosen such that the initial \gls{CSO} is maximized (see the Supplemental Material~\cite{supp_mat} for the details).
        Inset: Asymptotic value of the~\gls{CSO} for $t \to \infty$, indicating the necessity to perform the quench inside the disordered phase of the~\gls{TFIM}.
    }
    \label{fig:coherences}
\end{figure}

Quantum mechanics is characterized by a tension between continuous deterministic processes and discontinuous stochastic ones.
According to the common interpretation, a quantum system can exist in a coherent superposition of states evolving unitarily under the Schrödinger equation until it interacts with a measurement apparatus, which collapses the superposition to a specific state following Born’s rule~\cite{Neumann1955}.
The puzzling, sudden emergence of an objective outcome from a superposition of possible ones is known as the measurement problem~\cite{Wigner1963,Schlosshauer2005, Brukner2015, Tomaz2025}.
Attempts to solve this problem led to the development of alternatives to the Copenhagen interpretation, such as the many\hyp worlds~\cite{Everett1956} and the relational interpretation~\cite{Rovelli1996} as well as information\hyp theoretic approaches~\cite{Brukner2003}. 
Moreover, modifications of the Schrödinger equation have been proposed, the most notable being objective collapse models~\cite{Ghirardi1986, Bassi2003, Penrose1996, Oppenheim2023} and Bohmian mechanics~\cite{Bohm1952, Durr2009}.
A different line of thought that has been followed in the context of the measurement problem is labeled the \textit{decoherence} program~\cite{Zeh1970, Zeh1973, Zurek2003}.
This approach is solely based on standard quantum mechanics and maintains that the superpositions of a quantum system are quickly suppressed when it interacts with an environment due to the buildup of quantum correlations.
While successful, there is consensus that decoherence, in general, only solves the pre\hyp measurement problem, i.e., it explains the transition from quantum coherent to incoherent probability distributions, but it says little about how a collapse onto a particular state takes place.
Despite experimental evidence consistent with nonlocal, effectively instantaneous quantum collapses~\cite{Fuwa2015, Garrisi2019}, recent theoretical analyses have challenged the underlying assumptions of the measurement postulate.
Notably, bounds on the times necessary to implement pre\hyp measurements in isolated systems~\cite{Strasberg2022} and full measurements in open systems~\cite{Shettell2023} have been derived.
Moreover, it has been rigorously shown that ideal projective measurements require infinite thermodynamic resources~\cite{Guryanova2020}, while the accuracy of approximate measurements can be exponentially enhanced by increasing the size of the observing system~\cite{Schwarzhans2023}.
In this letter, we address the measurement problem from a different angle.
We propose an experimental test using modern analog quantum simulators, which can realize fast, repeated measurements while allowing for precise control of the Hamiltonian parameters and long coherence times.
Our idea is based on the intuition that if the projective, instantaneous description of measurements only constitutes an approximation to the actual physical measurement process, then small deviations in the system's dynamics, caused by the incomplete description of the measurement process, should be amplified under repeated measurements.
This statement can be formulated in a precise manner by studying the statistics of repeated measurement outcomes of observables, following a global quantum quench, shown schematically in \cref{fig:first:figure}.
For that purpose, we analyze the dynamics of (local) observables subject to repeated, projective measurements, and the probability to confirm previous measurement outcomes.
The system under consideration is the~\acrfull{TFIM}, which, under certain assumptions on the quench protocol, drives an initially pure subsystem to a fully incoherent classical mixture of distinct measurement outcomes in the long\hyp time limit.
In order to understand whether violations of the projective measurement assumption yield practically detectable deviations, we introduce an alternative, continuous formulation, in which measurements only cause a classical reduction of the amplitudes of a quantum state.
Under mild assumptions, this alternative approach to the measurement description yields significantly different statistics for repeated measurement outcomes than the projective one.
We argue that this discrepancy is independent of our specific proposal: Predictions due to any continuous formulation of quantum measurements are in conflict with the results obtained from standard theory.
We stress that we do not aim to address the measurement problem in its full complexity~\cite{Hance2022,Muller2023}, but only to investigate the implications of the description of measurements as instantaneous projections in a high\hyp frequency repeated\hyp measurement scenario.
\paragraph{Projective measurements-- }
We start from the textbook perspective on measurements, that is, regarding them as instantaneous, irreversible and non\hyp deterministic processes causing wave\hyp function collapses.
When describing quantum measurements, it is crucial to incorporate the inevitable influence of a measured system's environment.
While different methods have been proposed to model these effects~\cite{Schlosshauer2019}, we include the environment as part of the considered system.
This explicit treatment allows for precise control over the environmental degrees of freedom, which are widely believed to play a central role in the context of measurements~\cite{Zurek1981, Zurek2003, Joos2003}.
We consider the~\gls{TFIM}
\begin{align}
    \hat H = - J \sum_{j} \hat\sigma_j^x \hat\sigma_{j+1}^x + h \sum_{j} \hat \sigma_j^z \,,
    \label{eq:TFIM:Hamiltonian}
\end{align}
where each site $j$ corresponds to a spin\hyp $1/2$ degree of freedom and Pauli\hyp matrices are denoted as $\hat{\sigma}_j^{x,y,z}$.
The ratio of spin\hyp interaction strength $J$ and transverse field $h$ fully captures the physics of the \gls{TFIM} and the system features a quantum\hyp critical point for $\xi \equiv h/J = 1$, separating ordered, $\xi < 1$, and disordered phases, $\xi > 1$.
For measurements on an $N$\hyp sited subsystem in the thermodynamic limit, the remaining spins constitute an environment, which is coupled to the subsystem only via two boundary spins.
Choosing this coupling large enough, the influence of further external environments can be neglected~\cite{Gross2017,Bernien2017}.
We initialize the non\hyp equilibrium dynamics with a global product state, so that the reduced state on the N\hyp sited subsystem is pure and the effect of projective measurements is most pronounced.
In detail, we prepare the \gls{TFIM} in the ground state for an initial field $h_0 \to \infty$ and quench instantaneously to $h \in \mathbb{R}_0^+$ at time $t = 0$.
We study the probability of confirming a measurement outcome under repeated measurements.
In the simplest case, one could choose to measure a conserved quantity, which projects the state into the subspace corresponding to the measurement result, where it remains for all times.
Thus, in principle, any differing value measured in subsequent observations would violate the theory of projective measurements.
The \gls{TFIM}, however, does not exhibit conserved quantities that are defined on a finite subsystem only.
We thus consider observables maximizing the time $t_\mathrm{max}$ between two measurements, such that the measurement outcome is confirmed with a probability of at least $P_\mathrm{min}$, which we refer to as \textit{quasi\hyp conserved} quantities.
Unsurprisingly, these are precisely the conserved quantities $\hat\Pi_\nu$ of the isolated open chain of length $N$,  with $\hat\Pi_\nu\ket{\Pi_\nu} = \Pi_\nu\ket{\Pi_\nu}$ ($\Pi_\nu \in \left\{0,1\right\}$).
As we show in the Supplemental Material~\cite{supp_mat}, measuring the occupation of mode $\nu$ yields
\begin{align} \label{eq:projective_t_max}
    t_\mathrm{max} \geq t_\mathrm{bound} \propto \frac{1 - P_\mathrm{min}}{\mathcal{N}_\nu J} \,.
\end{align}
The normalization constant $\mathcal{N}_\nu \propto \sqrt{1 / N}$ manifests the control over $t_\mathrm{max}$ by tuning the subsystem size and thereby reaching experimentally accessible inter\hyp measurement timescales.
This scaling is robust with respect to longer\hyp ranged interactions but highly dependent on dimensionality, where the experimentally favorable relation $t_\mathrm{bound} \propto \sqrt{N}$ only holds in one dimension.
From now on, we will drop the label $\nu$, since our results apply to all eigenmodes.
Considering $n$ repeated measurements, we define the relative incidence $w_n = \sum_{j=1}^n (1 - |\Pi(t_j) - \Pi(t_0)|) / n$ of confirming the initially measured mode occupation $\Pi(t_0)$ when performing the $j^\mathrm{th}$ measurement at time $t_j$.
According to \cref{eq:projective_t_max}, if $t_j - t_{j-1} \leq t_\mathrm{bound}$, the probability of the event $w_n = 1$ can be bounded by $\mathcal{P}_\mathrm{proj}(w_n = 1) \geq (P_\mathrm{min})^n \equiv \mathcal{P}_\mathrm{proj}^\mathrm{min}(w_n=1)$ for projective measurements and we refer to the series of $\mathcal{P}_X(w_n = 1)$ observed at different times $t_j$ as the~\acrlong{RMS} (here $X=\mathrm{proj}$, and $X=\mathrm{dec}$ for the case of continuous measurements discussed below).
In \cref{fig:measurement_statistics}, upper and lower bounds on the~\acrfull{RMS} are shown for the case of $N=8$ spins.
Note the high probability to confirm the initial measurement result, which decays only very slowly as a function of time, with the number of confirmations.
We emphasize that for the chosen parameters, the inter\hyp measurement timescale is of the order of $1\mathrm{ms}$, if the protocol is realized on an optical lattice~\cite{Meinert2013}.
The suppression of different measurement results due to stroboscopically repeated observations is reminiscent of quantum Zeno\hyp type experiments~\cite{Misra1977, Schafer2014, Kalb2016}.
However, we stress that here we focus on the repeated measurement of quasi\hyp conserved quantities. 
In the limiting case $N\to \infty$, these become exactly conserved, and thus a single measurement would suffice to freeze the dynamics entirely.
\paragraph{A Continuous Formulation-- }
There has been prominent support for the claim that wave\hyp function collapses are effective placeholders for an underlying physical process~\cite{Bell1990, Penrose1996}.
Arguably, decoherence has been one of the most widely discussed mechanisms in this context.
However, inasmuch as it originates from the problematic theory itself, decoherence could only shift the concerns rather than mitigating the problem.
Nevertheless, we will accept decoherence as a part of our continuous alternative to orthodox theory for two central reasons.
First, it allows us to develop a formulation without explicit wave\hyp function collapses that we show to be incompatible with standard theory.
It may therefore point to the components of quantum physics that actually \textit{need some attention}~\cite{Bell1990} in the sense that wave\hyp function collapses imply decoherence but not necessarily vice versa.
The second reason is that decoherence timescales are sufficiently small to explain why the projective formulation could have been \textit{just fine for all practical purposes}~\cite{Bell1990} so far.
Let us again consider the $N$\hyp sited open subsystem within the \gls{TFIM} and introduce a measure, which allows us to quantify decoherence, and which can be utilized to derive bounds for the~\gls{RMS} for a theory of non\hyp projective measurements.
The system's local properties are fully captured by the \gls{RDM} $\hat{\rho}_N(t)$.
While the \gls{TFIM} does not thermalize, it has been shown to equilibrate locally~\cite{Calabrese_2012_2}, with $\hat{\rho}_N(t)$ approaching the \gls{RDM} of a generalized Gibbs ensemble algebraically in time~\cite{Fagotti2013}.
We introduce the \gls{CSO} to quantify the overlap between the subspaces corresponding to our protocol's distinct measurement results $\Pi$ via
\begin{align} \label{eq:coherence_measure}
    \coherence{\hat{\rho}_N (t)}{} \equiv 2 \norm{(\mathbb{1} - \hat\Pi) \hat{\rho}_N(t) \hat\Pi}_\mathrm{Fr} \in [0, 1].
\end{align}
Here, $\norm{\cdot}_\mathrm{Fr}$ denotes the Frobenius norm (see the Supplemental Material~\cite{supp_mat} for further details).
Given the equilibration of the underlying state, $\coherence{\hat\rho_N(t)}{}$ converges $\propto t^{-3/2}$ as $t \to \infty$.
Note that following the same arguments employed for the scaling of $t_\mathrm{max}$ in the previous section, the decoherence timescale also increases with subsystem size $N$, which is shown in \cref{fig:coherences}.
We are interested in the scenario in which the time evolution completely suppresses \gls{CSO}, i.e., $\lim_{t \to \infty} \coherence{\hat\rho_N(t)}{} = 0$, such that decoherence fully accounts for the transition to a classical mixture of measurement outcomes.
As an interesting side remark, we found that this is achieved only for quenches which do not cross the quantum critical point, and that the initial \gls{CSO} is maximized precisely for quenches to $\xi = 1$~\cite{supp_mat}.
In~\cref{fig:coherences}, we illustrate the time dependence of the~\gls{CSO} for various subsystem sizes choosing $\xi=1$, while the infinite\hyp time value of the~\gls{CSO} is shown in the inset as a function of $\xi$, illustrating perfect decoherence as long as $\xi \geq 1$.
We now motivate our developed formalism for describing measurements as a continuous process with respect to the coherent part of the measured state, while collapsing the classical probabilities, only, and refer to the Supplemental Material~\cite{supp_mat} for an in\hyp depth analysis.
The main idea is to understand the observed measurement outcome as a realization of a classical random experiment.
We then use the fact that any reduced density operator of our subsystem can be expressed as a convex combination of an incoherent state $\hat \rho_\mathrm{ic}(t)$ with vanishing \gls{CSO} and a state $\hat \rho_\mathrm{c}(t)$ on the boundary of the set of density operators~\cite{Kimura2005}:
\begin{equation} \label{eq:state_split}
    \hat \rho(t) = \alpha (t) \hat \rho_\mathrm{c}(t) + (1-\alpha(t)) \hat \rho_\mathrm{ic}(t), ~ \alpha(t) \in [0,1] \; .
\end{equation}
A measurement at time $t$ is then described by a quantum channel, maintaining the contributions from  $\hat \rho_\mathrm{c}(t)$, while the incoherent part is subject to a projection into the subspace compatible with the measurement outcome $\Pi$:
\begin{equation} \label{eq:continuous_measurement_channel}
    \hat \rho(t) \mapsto \alpha(t) \hat \rho_\mathrm{c}(t) + (1 - \alpha(t)) \hat \Pi \hat \rho_\mathrm{ic}(t)\hat \Pi \; .
\end{equation}
Note that the decomposition \cref{eq:state_split} is somewhat arbitrary, yet once $\hat \rho_\mathrm{ic}(t)$ is fixed, the fact that any density operator can be decomposed as such a convex combination, in principle allows to evaluate the dynamics of the density operator under repeated measurements, and thus also the~\gls{CSO}.
Here, we suggest choosing $\hat \rho_\mathrm{ic}(t) \propto \mathbb 1$, 
i.e., the maximally mixed state on the space of reduced density operators for our given subsystem.
We justify this choice by noting that the maximally mixed state is \textit{the} unique state which exhibits no \gls{CSO} for \textit{any} measurement, thus avoiding any bias in the description of the measurement process, due to the chosen observable.
We remark that because $\hat \rho_\mathrm{c}(t)$ is a state on the boundary of the set of density operators, it cannot be expressed as a proper convex combination of the maximally mixed state and pure states, because boundary states are singular.
This is important since any pure state describes only coherent information and thus must not be affected, according to our proposed measurement postulate \cref{eq:continuous_measurement_channel}.
We can now evaluate upper bounds $\mathcal{P}_\mathrm{dec}^\mathrm{max}(w_n=1)$ for the probability $\mathcal{P}_\mathrm{dec}(w_n=1)$ to confirm previous measurement outcomes, which can be compared to the~\gls{RMS} obtained for projective measurements.
Choosing the same system configuration, the time evolution of $\mathcal{P}_\mathrm{dec}^\mathrm{max}(w_n=1)$ is shown in \cref{fig:measurement_statistics}, exhibiting a significantly faster decay, compared to the case of projective measurements.
The stark difference to the case of projective measurements can be understood by noting that repeated measurements periodically cancel a state's components within the subspace orthogonal to the measurement result due to wave\hyp function collapses.
Consequently, even small deviations compared to a formulation that does not assign a distinguished role to the act of observation are amplified with the number of measurements.
This behavior is thus qualitatively replicated by any collapse\hyp free theory and we therefore expect our findings to be independent of the specific choice of the theory.

\paragraph{Experimental realization-- }
Our protocol is based on measuring occupations of eigenmodes in the open subsystem.
The non\hyp locality of these modes may, however, present an obstacle to experimental realization.
Taking a quantum\hyp state tomographic approach, the desired information can be deduced from local properties in an ensemble of systems.
Among the most accessible observables, measuring spin\hyp $z$ projections for sites $j = 0, \dots, N-1$ can serve this purpose and we denote the corresponding lower bounds on repeated confirmations of measurement outcomes as $\tilde{\mathcal{P}}_\mathrm{proj}^\mathrm{min}(w_n=1)$, which are compared to the collapse\hyp free theory in \cref{fig:measurement_statistics}.
While this simplistic work\hyp around in principle still allows for a discrimination from the decoherence\hyp based results, only certain measurement outcomes are meaningful and low acceptance rates may hinder implementation in practice.
We refer to the Supplemental Material~\cite{supp_mat}, where we detail more sophisticated local measurement schemes that exhibit higher acceptance and feature larger deviations from the decoherence\hyp based protocol.
A necessary requirement for the implementation of our protocol is the ability to repeatedly alternate a measurement and a short time evolution for every sample  (the relevant timescale is specified in the caption of \cref{fig:measurement_statistics}).
For this, well\hyp suited platforms are, for instance, Rydberg atom arrays in optical tweezers~\cite{Borish2020,Scholl2021,Xu2024} and ultracold atoms in optical lattices~\cite{Bloch2012, Chen2011, Liao2021}, as well as digital quantum computing devices~\cite{Miessen2024, Acharya2025-wn, Haghshenas2025}.
Note that, compared to the idealized case considered here, experimental realizations will inevitably need to take into account errors stemming from an imperfect time evolution, noisy readout schemes and a finite number of samples.
We believe, however, that our protocol can be implemented relying on Trotterized dynamics due to the resetting effect of projective measurements, which would prevent amplification of the associated error.
Predictions according to collapse\hyp free theories, on the other hand, would neither be significantly affected by Trotter approximation since time evolution still induces transitions between the two subspaces discriminated by measurements in the open system.

\textit{Conclusions--}
We have analyzed the \gls{RMS} the one\hyp dimensional \gls{TFIM} in the thermodynamic limit with a finite number of adjacent spins representing the measured subsystem, using either the standard formulation of instantaneous, projective measurements, or a pure decoherence\hyp based approach.
For the latter case, we introduced a general model describing measurements as operations that maintain quantum coherences and reduce only the classical, incoherent part of the state, eliminating the necessity of an observer\hyp induced collapse.
Strict bounds for the relative incidences of confirming the first observation in subsequent measurements have been derived for both approaches.
We showed that the required inter-measurement timescales are controlled by the subsystems' linear dimension, reaching experimentally accessible frequencies for as few as $8$ spins.
Comparing the~\glspl{RMS} for collapse\hyp based and collapse\hyp free descriptions of measurements, we revealed significant deviations already after $\mathcal O(10)$ measurements.
This could allow for an experimental test of the validity of the concept of projective measurements in a simple system and using current quantum simulators.
Our theory of continuous measurements constitutes a pragmatic alternative to the standard approach; it incorporates decoherence while otherwise adhering as closely as possible to natural assumptions.
In fact, just as wave\hyp function collapses might be unphysical, decoherence could similarly provide only an alternative placeholder for the actual process underlying quantum measurements.
In this sense, we approached the measurement problem only on an epistemological rather than an ontological level.
However, the strong deviations between the derived~\gls{RMS} manifest incompatibility of physical realities for which projective measurements and our continuous formulation, respectively, provide effective descriptions.
We emphasize that the key ingredient for this conclusion is solely the absence of explicit wave\hyp function collapses and thus not restricted to our specific decoherence\hyp based proposal.
Consequently, this incompatibility persists between standard theory and any reformulation yielding continuous measurement processes.
Leveraging modern quantum simulation platforms capable of high-frequency measurements, we seek to address the conflict through empirical investigation, with an approach akin to Bell’s theorem and the Freedman\hyp Clauser experiment in the debate over local hidden\hyp variable theories.

\paragraph{Acknowledgements--}
We are grateful to Reja Wilke for useful comments on a previous version of the manuscript.
MM acknowledges funding from the Royal Society and Research Ireland.
SP acknowledges support by the Deutsche Forschungsgemeinschaft (DFG, German Research Foundation) under Germany’s Excellence Strategy-426 EXC-2111-390814868.
This work was supported by Grant No. INST 86/1885-1 FUGG of the German Research Foundation (DFG).
\bibliography{Literature}
%
\appendix
\setcounter{section}{1}
\setcounter{equation}{0}%
\setcounter{figure}{0}%
\setcounter{table}{0}%
\makeatletter%
\renewcommand{\thesection}{S\arabic{section}}%
\renewcommand{\thesubsection}{\thesection.\arabic{subsection}}%
\renewcommand{\theequation}{S\arabic{equation}}%
\renewcommand{\thefigure}{S\arabic{figure}}%
\renewcommand{\bibnumfmt}[1]{[S#1]}%
\makeatother
%
\widetext
%
\begin{center}
\textbf{\large Supplemental Material for \\``Probing the Physical Reality of Projective Measurements''}
\end{center}
\numberwithin{equation}{section}
\onecolumngrid
\section{The \acrlong{TFIM}}
We consider the one\hyp dimensional \acrfull{TFIM}
\begin{align} \label{eq:sm:ham_spin_tfim}
   \hat H = - J \sum_{j=0}^{L-\nu} \hat\sigma_j^x \hat\sigma_{j+1}^x + h \sum_{j=0}^{L-1} \hat\sigma_j^z,
\end{align}
where we denoted the interaction strength by $J$, the transverse field by $h$, and $\hat\sigma_j^{\alpha}$ represent Pauli operators on the single\hyp spin space $\mathbb{C}^2$ of site $j$ in a chain of $L$ sites.
For $\nu = 2$ the system exhibits \glspl{OBC}, while for \glspl{PBC} $\nu = 1$ applies and periodicity is introduced by defining $\hat\sigma_{L}^\alpha \equiv \hat\sigma_0^\alpha$.
While the solution strategies for the two types of boundary conditions differ, they share the common starting point of transforming the spin degrees of freedom to spinless fermions in terms of a \gls{JW} transformation
\begin{align} \label{eq:sm:jw_trafo}
    \hat\sigma_j^x = \exp\left( i \pi \sum_{l<j} \hat c_l^\dagger \hat c_l^\nodagger \right) \left( \hat c_j^\dagger + \hat c_j^\nodagger \right),~
    \hat\sigma_j^y = - i \exp\left( i \pi \sum_{l<j} \hat c_l^\dagger \hat c_l^\nodagger \right) \left(\hat c_j^\dagger - \hat c_j^\nodagger \right),~
    \hat\sigma_j^z = 2 \hat c_j^\dagger \hat c_j^\nodagger - 1,
\end{align}
obeying the anticommutation relations $\{\hat c_j^\dagger,\hat c_l^\nodagger \} = \delta_{jl}^\nodagger$ and $\{\hat c_j^{(\dagger)}, \hat c_l^{(\dagger)} \} = 0$.
Here and throughout, sums of some index $j$ over $j<l$ are indexed beginning with $0$, i.e. $j=0,...,l-1$.
\subsection{The periodic \acrlong{TFIM}}
For \glspl{PBC}, the model can be solved analytically \cite{Pfeuty1970} and the ground state is found in the even fermion\hyp parity subspace, called \gls{NS} sector.
Assuming $L \in 2 \mathbb{N}$ and setting the ground state energy to zero we obtain
\begin{align}
    \hat H_\text{NS} = J \sum_{k \in K_\text{NS}} \epsilon_k(\xi) \hat\gamma_k^\dagger \hat\gamma_k^\nodagger,~ \epsilon_k(\xi) \equiv 2 \sqrt{1 + \xi^2 - 2 \xi \cos k}
\end{align}
where $K_\text{NS} \equiv \{ (2\pi n + 1/2) / L ~|~ n = -L/2, ..., L/2 - 1 \}$ and $\hat{\gamma}_k^{(\dagger)}$ describe Bogoliubov fermions.
The ground state in the \gls{NS} subspace and thus the global ground state is the vacuum of Bogoliubov fermions, $\hat\gamma_k^\nodagger \ket{\varnothing (\xi)} = 0 ~ \forall k \in K_\text{NS}$.
The Bogoliubov angle $\theta_k(\xi)$ is defined by
\begin{align}
    \exp(i \theta_k(\xi)) = \frac{\xi - \exp(ik)}{\sqrt{1 + \xi^2 - 2 \xi \cos k}}.
\end{align}
\subsection{The open \acrlong{TFIM}} \label{sec:open_tfim}
For the sake of consistency in notation, we shall denote the number of sites in the open chain as $N$ and the corresponding Hamiltonian $H_N$ for the \gls{TFIM} with \glspl{OBC} here onward.
Closely following Refs.~\cite{He2017, Lieb1961}, broken translation symmetry suggests proceeding directly to a Bogoliubov transformation in position representation,
\begin{align} \label{eq:sm:tfim_obc_modes}
    \hat\eta_k \equiv \sum_{j=0}^{N-1} (g_{kj}^\nodagger \hat c_j^\nodagger + h_{kj}^\nodagger \hat c_j^\dagger),~ k \in \{ 0, ..., N-1 \}.
\end{align}
Demanding $\hat\eta_k^{(\dagger)}$ to diagonalize the open chain yields Bogoliubov mode energies $\lambda_k = J [4 (1 + \xi^2) - 8\xi \cos \varphi_k]$ as well as
\begin{align} \label{eq:sm:tfim_obc_mode_gh}
    g_{kj} = \frac{\mathcal{N}_k}{2} ( \sin{((N-j) \varphi_k)} + \sin{((j+1) \varphi_k)} ),~ h_{kj} = \frac{\mathcal{N}_k}{2} ( \sin{((N-j) \varphi_k)} - \sin{((j+1) \varphi_k)} ).
\end{align}
Normalization is ensured by $\mathcal{N}_k \equiv (\sum_j \sin^2 ((j+1) \varphi_k))^{-1/2}$ and the angles $\{ \varphi_k \in (0, \pi) | k = 0, ..., N-1 \}$ are the solutions, defined modulo $\pi$, of the transcendental equation
\begin{align} \label{eq:sm:obc_transcendental}
    \frac{\sin{(N \varphi_k)}}{\sin{((N+1) \varphi_k)}} = \xi,
\end{align}
where we define the ordering $\varphi_k < \varphi_{k+1}$.
Notably, this equation only yields the complete set of eigenmodes as long as $N / (N+1) < |\xi|$.
However, the regime relevant to our work is contained within the parameter range for which \cref{eq:sm:obc_transcendental} determines all eigenmodes.

We note here that in Fig. 2 in the main text, mode $k=1$ was considered. In Fig. 3 in the main text, mode $k=0$ was considered for $N=2$, while we used $k=1$ for $N=4, 8, 16$.
\section{Quantum quenches}
Quantum quenches refer the procedure of preparing a system in the ground state of some Hamiltonian and subsequently evolving it in time with a different Hamiltonian, quenched to instantaneously.
For the \gls{TFIM} with \glspl{PBC} this corresponds to the evolution of $\ket{\varnothing (\xi_0)}$ under the the Hamiltonian $\hat H$ for the quenched\hyp to parameter $\xi$.
We consider the quench from the classical model, $\xi_0 \to \infty$, of non-interacting spins, for which the non-degenerate ground state can be trivially determined in terms of spins as a fully z\hyp polarized chain, i.e. $\ket{\Psi(t \leq 0)} = \ket{\downarrow} \otimes \cdots \otimes \ket{\downarrow}$, to some finite parameter ratio $\xi$ at time $t=0$.
For this scenario, Bogliubov modes coincide with the \gls{JW} fermions for $t \leq 0$, $\hat c_k^{(\dagger)} = \hat \gamma_k^{(\dagger)} |_{t \leq 0}$, and we can express the pure state of the entire chain under \glspl{PBC} evolving in time as
\begin{align} \label{eq:sm:quench_state}
    \ket{\Psi (t > 0)} = \exp(-i\hat H t) \ket{0} = \prod_{0 < k \in K_\text{NS}} \left( \cos{\frac{\theta_{k}(\xi)}{2}} + i \sin{\frac{\theta_k(\xi)}{2} \exp(-2i \epsilon_k(\xi) t) \hat\gamma_{-k}^\dagger \hat\gamma_k^\dagger} \right) \ket{\varnothing(\xi)},
\end{align}
with $\ket{0} = \lim_{\xi_0 \to \infty} \ket{\varnothing(\xi_0)}$, i.e $\hat c_j \ket{0} = 0 ~ \forall j$.
Importantly, this choice of $\xi_0$ implies the initial state to be of product form and, thus, when dividing the spin chain in any two blocks, there is no entanglement among the two open subsystems.
The measurement of an observable on one of the two blocks for which the state of the second block evolves to orthogonal subspaces for distinct measurement outcomes leads to \textit{full} decoherence of the subspace\hyp overlap between different measurement results.
By this, we refer to the transition of a pure superposition of distinct measurement outcomes to a completely classical mixture of associated probabilities over time due to tracing out part of the system.
The proposed protocol relates to experiments via the preparation of the chain for $t<0$ under the influence of a strong magnetic field $h_0 \gg J$, which is reduced in strength to $h = J \xi$ at $t=0$ much faster than the relevant timescale set by $\epsilon_\pi (\xi) = 2 \abs{1 + \xi}$.
\section{\Acrlong{RDM}}
Sites $0$ to $N-1$ constitute a subsystem embedded into the \gls{TFIM} with \glspl{PBC} of length $L>N$, interacting with the remaining chain via the coupling of adjacent spins on sites $L-1$ and $0$ as well as $N-1$ and $N$.
The subsystem is fully described in terms of the \acrfull{RDM} $\hat\rho_N (t) = \Tr_{L \backslash N} \ketbra{\Psi (t)}{\Psi (t)}$, where the index $L \backslash N$ indicates tracing out spins $N$ to $L-1$.
It has been shown that the non\hyp equilibrium dynamics in \cref{eq:sm:quench_state} describe local relaxation, meaning that the infinite time limit of $\hat\rho_N (t)$ is equal to the partial trace of a generalized Gibbs ensemble describing a mixed state of the full chain \cite{Calabrese_2012_2}.
In particular, $\hat\rho_N (t)$ is guaranteed to approach a stationary state for $t \to \infty$.
While it has further been shown that the latter happens $\propto t^{-3/2}$ \cite{Fagotti2013}, we have to resolve the short\hyp time dynamics for studying the $N$\hyp site subsystem decohere into the environment modelled by the remaining spins.
We introduce Majorana fermions defined by
\begin{align}
    \hat a_{2l}^\nodagger \equiv \left( \prod_{j<l} \hat\sigma_j^z \right) \hat\sigma_l^y = i \left( \hat c_l^\nodagger - \hat c_l^\dagger \right),~
    \hat a_{2l+1}^\nodagger \equiv \left( \prod_{j<l} \hat\sigma_j^z \right) \hat\sigma_l^x = \hat c_l^\nodagger + \hat c_l^\dagger,
\end{align}
fulfilling $\hat a_j^\dagger = \hat a_j^\nodagger$ and $\{ \hat a_j^\nodagger, \hat a_l^\nodagger \} = 2 \delta_{jl}$.
Wick's theorem \cite{Molinari2023} applies such that the $2N \times 2N$ correlation matrix $\Gamma_{jl} (t) \equiv \Tr (\hat\rho_N (t) \hat a_l \hat a_j) - \delta_{jl}$ entails all information about the subsystem and the \gls{RDM} becomes Gaussian~\cite{Peschel2003},
\begin{align} \label{eq:sm:rdm_exponential}
    \hat\rho_N (t) = \frac{1}{Z} \exp \left( \frac{1}{4} \sum_{j,l} \hat a_j [W(t)]_{j,l} \hat a_l \right).
\end{align}
The pre\hyp factor ensures normalization and $\tanh (W(t)/2) = \Gamma (t)$ holds true but we avoid dealing with $W$ explicitly.
$\Gamma (t)$ is of the form
\begin{align} \label{eq:sm:rdm_correlation}
    \Gamma (t) = 
    {\renewcommand{\arraystretch}{1.5}
    \begin{pmatrix}
        \Gamma_0 & \Gamma_{-1} & \cdots & \Gamma_{1-N} \\
        \Gamma_1 & \Gamma_{0} & \cdots & \Gamma_{2-N} \\
        \vdots & \vdots & \ddots & \vdots \\
        \Gamma_{N-1} & \Gamma_{N-1} & \cdots & \Gamma_0
    \end{pmatrix}
    }.
\end{align}
We introduce the thermodynamic limit $L \to \infty$ at this point for which the constituting $2 \times 2$ blocks read
\begin{align}
    \Gamma_l = \frac{i}{\pi} \int\limits_0^\pi \mathrm{d}k
    {\renewcommand{\arraystretch}{1.5}
    \begin{pmatrix}
        - f(k) & g_-(k) \\
        g_+(k) & f(k)
    \end{pmatrix}
    },
\end{align}
defining
\begin{equation}
\begin{aligned}
    f (k) &= \sin (kl) \sin \theta_k(\xi) \sin (2 \epsilon_k(\xi) t) \\
    g_\pm (k) &= \pm [ \cos \theta_k (\xi) \cos (\theta_k (\xi) \mp kl) + \cos (2 \epsilon_k (\xi) t) \sin \theta_k (\xi) \sin (\theta_k(\xi) \mp kl) ].
\end{aligned}
\end{equation}
The aforementioned relaxation $\propto  t^{-3/2}$ of $\hat \rho_N (t)$ implies the same scaling for the Majorana correlation matrix approaching equilibrium element\hyp wise.
\section{Projective measurements} \label{ch:sm:projective_measurements}
We take on the textbook perspective on quantum measurements and consider the $N$\hyp sited subsystem.
Based on the result we obtained in a first measurement, we want to make predictions on the following ones.
There are, however, no conserved quantities in the \gls{TFIM} that can be observed locally on a finite number of sites.
Thus, we want to maximize the probability $\hat P_\mathrm{proj} (t)$ of confirming the measurement result observed at time $t_0$ after some time $t \geq 0$
\begin{align} \label{eq:sm:proj_measurement_prob}
    \hat P_\mathrm{proj} (t) = \frac{\bra{\Psi(t_0)} \hat \Pi \exp(i\hat Ht) \hat\Pi \exp(-i\hat Ht) \hat\Pi \ket{\Psi(t_0)}}{\bra{\Psi(t_0)} \hat\Pi \ket{\Psi(t_0)}} = 1 - \frac{\bra{\Psi(t_0)} \hat\Pi \exp(i\hat Ht) [\exp(-i\hat Ht), \hat\Pi] \hat\Pi \ket{\Psi(t_0)}}{\norm{\hat\Pi \ket{\Psi(t_0)}}^2},
\end{align}
where $\hat\Pi$ projects onto the subspace associated with measurement result $1$.
In full generality, we assume that one wants to confirm a measured value of $1$ but the reasoning is equally valid for a first measurement result of $0$ by replacing $\hat\Pi$ by $\mathbb{1} - \hat\Pi$.
The second summand on the right-hand side of \cref{eq:sm:proj_measurement_prob} can be bounded from above using the operator norm and its submultiplicativity, $\hat P_\mathrm{proj} (t) \geq 1 - \norm{[\hat\Pi, \exp\small(-i\hat Ht \small)]}_\mathrm{Op}$, due to its Hermiticity.
We use the identity
\begin{align}
    \left[ \hat\Pi, \exp(-i\hat Ht) \right] = -it \int\limits_0^1 \mathrm{d}s \exp(-i(1-s)\hat Ht) [\hat \Pi, \hat H] \exp(-is\hat Ht),
\end{align}
by taking its operator norm and employing the triangle inequality for integrals, submultiplicativity of the operator norm and the fact that the operator norm of a unitary operator equals $1$ yielding
\begin{align} \label{eq:sm:proj_measurement_bound}
    \hat P_\mathrm{proj} (t) \geq 1 - t \norm{\left[\hat\Pi, \hat H \right]}_\mathrm{Op}.
\end{align}
Notably, the same bound is obtained to first order in time by expanding $\exp \small(\pm i \hat Ht\small)$ in \cref{eq:sm:proj_measurement_prob} and we thus argue to capture the essential scaling for the relevant small timescales and stress having shown that our bound holds exactly.
\newline \indent
Let us decompose the Hamiltonian as $\hat H = \hat H_N + \hat H_{L \backslash N} + H_\mathrm{int}$ into individual terms for the finite subsystem under \glspl{OBC}, the remaining spins and the interaction between the two constituents of the full chain, respectively.
This implies that only the last term is relevant for the bound \cref{eq:sm:proj_measurement_bound} if we choose $\hat\Pi$ to project onto fixed occupancies of the eigenmodes \cref{eq:sm:tfim_obc_modes} of the $N$\hyp site \glspl{OBC} \gls{TFIM}, i.e. $\hat\Pi_k = \hat\eta_k^\dagger \hat\eta_k^\nodagger$.
Consequently,
\begin{align}
    \norm{\left[ \hat H, \hat\eta_k^\dagger \hat\eta_k^\nodagger \right]}_\mathrm{Op} = \sqrt{2} J \mathcal{N}_k \abs{\sin (N \varphi_k)},
\end{align}
where $\mathcal{N}_k \sim \sqrt{2/N}$ as $N \to \infty$.
The answer to the question of what is the maximal time $t_\mathrm{max}$ to wait after a projective measurement of $\hat\Pi_k$ and still being guaranteed to obtain the same result with probability $P_\mathrm{proj} (t_\mathrm{max}) \geq P_\mathrm{min}$ upon measuring again is therefore bounded by
\begin{align} \label{eq:sm:t_max_bound}
    t_\mathrm{max} \geq t_\mathrm{bound} \equiv \frac{1-P_\mathrm{min}}{\sqrt{2} \mathcal{N}_k J \abs{\sin (N \varphi_k)}} \in \mathcal{O} (N^{1/2}).
\end{align}
The emphasis is on the control of this time scale via the scaling $\sim \sqrt{N}$, which is recovered for any finite\hyp range interaction in one dimension while a $d$\hyp dimensional system exhibits timescales $\sim N^{(2 - d)/2}$.
\cref{eq:sm:t_max_bound} clearly yields practically accessible bounds for the measurement timescales: for $N=8$ spins, $\xi = 1$ and $P_\mathrm{min} = 0.99$ such that the probability for confirming the first measurement outcome ten consecutive times is still $(P_\mathrm{proj} (t \leq t_\mathrm{max}))^{10} \geq (P_\mathrm{min})^{10} > 0.9$, a bound of $t_\mathrm{bound} > 0.015 / J$ is reached.
Coupling strengths of $J = 25\, \mathrm{Hz}$ have already been realized in experimental setups of quench protocols in the \gls{TFIM} over a decade ago \cite{Meinert2013} and imply a bound of $t_\mathrm{bound} > 600\, \mu \mathrm{s}$.
\section{Decoherence} \label{ch:sm:decoherence}
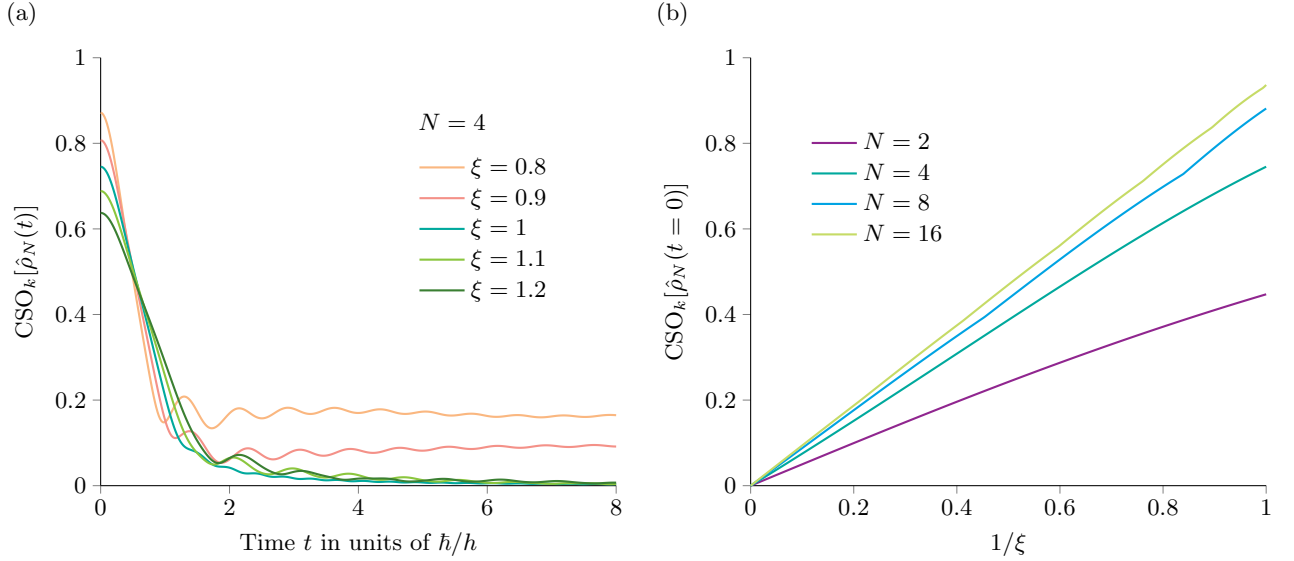
\begin{figure}
    \centering
    \begin{minipage}{0.45\textwidth}
        \centering
        \begin{tikzpicture}
            \begin{axis}[
                ymin=0, ymax=1,
                xmin=0, xmax=8,
                axis y line=left,
                xtick style={/pgfplots/major tick length=3pt},
                xtick style={yshift=-1.5pt},
                ytick style={/pgfplots/major tick length=3pt},
                ytick style={xshift=-1.5pt},
                xlabel={Time $t$ in units of $\hbar/h$},
                ylabel={$\coherence{\hat\rho_N (t)}{k}$},
                legend cell align=left,
                legend style={
                    draw=none,
                    at={(0.6,0.6)},
                    anchor=west
                },
                axis x line=bottom,
                axis line style={-},
                clip=false
            ]
                \node[anchor=west] at (rel axis cs: 0.6, 0.85) {$N=4$};
                \node[anchor=north west] at (rel axis cs:-0.2,1.15) {(a)};
        
                \addplot [
                    solid,
                    mark=none,
                    thick,
                    draw=Apricot
                ] table [
                    x index=0,
                    y index=1,
                    col sep=space
                ] {data/coherence_n_4_xi_0p8.dat};
                \addlegendentry{$\xi = 0.8$}
            
                \addplot [
                    solid,
                    mark=none,
                    thick,
                    draw=Salmon
                ] table [
                    x index=0,
                    y index=1,
                    col sep=space
                ] {data/coherence_n_4_xi_0p9.dat};
                \addlegendentry{$\xi = 0.9$}
            
                \addplot [
                    solid,
                    mark=none,
                    thick,
                    draw=Emerald
                ] table [
                    x index=0,
                    y index=1,
                    col sep=space
                ] {data/coherence_n_4_xi_1.dat};
                \addlegendentry{$\xi = 1$}
            
                \addplot [
                    solid,
                    mark=none,
                    thick,
                    draw=LimeGreen
                ] table [
                    x index=0,
                    y index=1,
                    col sep=space
                ] {data/coherence_n_4_xi_1p1.dat};
                \addlegendentry{$\xi = 1.1$}
        
                \addplot [
                    solid,
                    mark=none,
                    thick,
                    draw=OliveGreen
                ] table [
                    x index=0,
                    y index=1,
                    col sep=space
                ] {data/coherence_n_4_xi_1p2.dat};
                \addlegendentry{$\xi = 1.2$}
            \end{axis}
        \end{tikzpicture}
    \end{minipage}
    \hspace{0.02\textwidth}
    \begin{minipage}{0.45\textwidth}
        \centering
        \begin{tikzpicture}
            \begin{axis}[
                ymin=0, ymax=1,
                xmin=0, xmax=1,
                axis y line=left,
                xtick style={/pgfplots/major tick length=3pt},
                xtick style={yshift=-1.5pt},
                ytick style={/pgfplots/major tick length=3pt},
                ytick style={xshift=-1.5pt},
                xlabel={$1/\xi$},
                ylabel={$\coherence{\hat\rho_N (t=0)}{k}$},
                legend cell align=left,
                legend style={
                    draw=none,
                    at={(0.1,0.85)},
                    anchor=north west
                },
                axis x line=bottom,
                axis line style={-},
                clip=false
            ]
                \node[anchor=north west] at (rel axis cs:-0.2,1.15) {(b)};
        
                \addplot [
                    solid,
                    mark=none,
                    thick,
                    draw=Plum
                ] table [
                    x index=0,
                    y index=1,
                    col sep=space
                ] {data/initial_coherences.dat};
                \addlegendentry{$N = 2$}
            
                \addplot [
                    solid,
                    mark=none,
                    thick,
                    draw=Emerald
                ] table [
                    x index=0,
                    y index=2,
                    col sep=space
                ] {data/initial_coherences.dat};
                \addlegendentry{$N = 4$}
            
                \addplot [
                    solid,
                    mark=none,
                    thick,
                    draw=Cerulean
                ] table [
                    x index=0,
                    y index=3,
                    col sep=space
                ] {data/initial_coherences.dat};
                \addlegendentry{$N = 8$}
            
                \addplot [
                    solid,
                    mark=none,
                    thick,
                    draw=SpringGreen
                ] table [
                    x index=0,
                    y index=4,
                    col sep=space
                ] {data/initial_coherences.dat};
                \addlegendentry{$N = 16$}
            \end{axis}
        \end{tikzpicture}
    \end{minipage}
    \caption{
    (a) Time dependence of \gls{CSO} for different quenched\hyp to values of $\xi$.
    The quench originates in the disordered phase at $\xi_0 \to \infty$.
    Note that $\xi = 1$ corresponds to the quantum critical point.
    (b) Initial \gls{CSO} for different system sizes $N$ within the disordered phase.
    The modes $k$ have been optimized for maximal \gls{CSO}; $k=1$ in (a) while $k$ is non-constant with $1/\xi$ in (b).
    }
    \label{sm:fig:initial_and_limiting_coherence}
\end{figure}

Decoherence describes a system's state transitioning from a superposition to a mixed state due to time evolution introducing entanglement with its environment.
While the role of decoherence in the context of the measurement problem has been debated since the concept was established, it is clearly a quantum mechanical process independent of interpretation and does describe a quantum\hyp to\hyp classical transition for suitably chosen observables.
If, however, projective measurements are ruled out in favor of decoherence, there are two central problems that arise:
\begin{enumerate*}[label=(\roman*)]
    \item the emergence of probabilities has to be supplied to the theory after all, specifically Born's rule \cite{Schlosshauer2019}, and
    \item the environmental selection of the measured system's states, which constitute the classical mixture eventually, has to be specified.
\end{enumerate*}
While proposals addressing the second issue started with work by Zurek \cite{Zurek1981, Zurek1982}, who coined the term \textit{ein\hyp selection}, it is the first issue that is the main concern and decoherence provides no answer to, thus preventing it from solving the measurement problem~\cite{Schlosshauer2019}.
However, if the above issues can be addressed and time evolution for quantum systems can be formulated in a unified and continuous manner, for example, Schrödinger's equation describing it all, while a \textit{human} or \textit{conscious observer} plays no extraordinary role, decoherence may provide a framework to describe quantum measurements.
It is not our aim to propose solutions to either issue (i) or (ii) but instead we approach the problem by asking what the implications of the occurrence and absence of wave function collapses are.
With decoherence typically taking place at extremely short timescales, projective measurements may have been \textit{just fine for all practical purposes} (John Bell, \cite{Bell1990}) for a long time, but as experimentally resolvable timescales have been greatly improved upon since the advent of quantum theory, we aim to make the difference of decoherence and collapses observable in the \gls{TFIM}.
\newline \indent
In order to quantify timescales, we have to define a state's \gls{CSO} among subspaces corresponding to distinct measurement results in the first place.
Clearly, this depends on the observable measured.
Connecting to the previous section, we proceed for measurements of $\hat\Pi_k = \hat\eta_k^\dagger \hat\eta_k$ and thus have to quantify \gls{CSO} between the subspaces $\Pi_k = 0$ and $\Pi_k = 1$, respectively.
We propose the Frobenius norm of the coherence matrix as the corresponding measure.
Notably, this choice does not provide a general measure of coherence~\cite{Baumgratz2014} but suffices here as we are only interested in the \gls{CSO}.
Hence, when representing the density operator $\hat\rho_N$ in the basis with well-defined occupation of the modes diagonalizing the isolated system as
\begin{align} \label{eq:sm:rdm_representation}
    \varrho_N = 
    {\renewcommand{\arraystretch}{1.5}
    \begin{pmatrix}
        \bra{0_k} \hat\rho_N \ket{0_k} & \bra{0_k} \hat\rho_N \ket{1_k} \\
        \bra{1_k} \hat\rho_N \ket{0_k} & \bra{1_k} \hat\rho_N \ket{1_k}
    \end{pmatrix},
    }
\end{align}
we quantify \gls{CSO} via the Frobenius norm of the off-diagonal $2^{N-1} \times 2^{N-1}$ dimensional blocks of the density matrix $\varrho_N$, corresponding to matrix elements between the sectors $0_k$ and $1_k$ with $\Pi_k = 0$ and $\Pi_k = 1$, respectively,
\begin{align} \label{eq:sm:coherence_measure}
    \coherence{\hat\rho_N (t)}{k} \equiv 2 \norm{ \bra{1_k} \hat\rho_N \ket{0_k} }_\mathrm{Fr}.
\end{align}
The prefactor implies $\coherence{\hat\rho_N}{k} \in [0, 1]$, where the greatest value, $\coherence{\hat\rho_N}{k} = 1$, is reached, for maximally entangled states between the two subspaces.
There are two central reasons advocating for this choice of $\coherence{\hat\rho_N}{k}$.
First of all, it is straightforward to show that this quantity is invariant under any unitary transformation $\hat U$ fulfilling $[\hat U, \hat\Pi_k] = 0$ and therefore captures the coherent overlap of the two eigenspaces independently of the representation within them.
The second reason concerns the semantics of $\coherence{\hat\rho_N (t)}{k}$.
It can be reformulated in terms of the purity of $\hat\rho_N$, $\mathrm{purity}(\hat\rho_N) = \Tr \hat\rho_N^2$, and a representation\hyp dependent manipulation of the same,
\begin{align} \label{eq:sm:cohernce_via_purity}
    \coherence{\hat\rho_N (t)}{k} = \sqrt{2} \left[
    \mathrm{purity}
    {\renewcommand{\arraystretch}{1.5}
    \begin{pmatrix}
        \bra{0_k} \hat\rho_N \ket{0_k} & \bra{0_k} \hat\rho_N \ket{1_k} \\
        \bra{1_k} \hat\rho_N \ket{0_k} & \bra{1_k} \hat\rho_N \ket{1_k}
    \end{pmatrix}
    }
    -
    \mathrm{purity}
    {\renewcommand{\arraystretch}{1.5}
    \begin{pmatrix}
        \bra{0_k} \hat\rho_N \ket{0_k} & 0 \\
        0 & \bra{1_k} \hat\rho_N \ket{1_k}
    \end{pmatrix}
    }
    \right]^{1/2}.
\end{align}
Thus, $(\coherence{\hat\rho_N (t)}{k})^2$ can be understood as a \textit{distance in purity} of $\varrho_N$ to the corresponding density matrix with fully incoherent subspace\hyp overlaps for the purpose of measuring $\hat\Pi_k$.
In particular, $\lim_{t \to \infty} \coherence{\hat\rho_N (t)}{k} = 0$ means that, for the given state and measurement, decoherence drives a quantum\hyp to\hyp classical transition.
It can be easily concluded from the formulation \cref{eq:sm:cohernce_via_purity} that $\coherence{\hat\rho_N (t)}{k}$ ignores the the mixedness of $\hat\Pi_k \hat\rho_N (t) \hat\Pi_k$ and $(\mathbb{1} - \hat\Pi_k) \hat\rho_N (t) (\mathbb{1} - \hat\Pi_k)$ and therefore only captures the coherent overlap between the $\Pi_k = 0, 1$ subspaces.
\newline \indent
Our \gls{CSO} measure can be computed efficiently as follows.
Starting from \cref{eq:sm:coherence_measure}, it is straightforward to derive
\begin{align} \label{eq:sm:coherence_trce_quead_exponentials}
\begin{aligned}
    \coherence{\hat\rho_N (t)}{k} &= \left[ 2 ~ \Tr_N \left( \left[ \hat\Pi_k \hat\rho_N \hat\Pi_k + (1 - \hat\Pi_k) \hat\rho_N (1-\hat\Pi_k) \right]^2 \right) \right]^{1/2} \\
    &= \frac{\sqrt{2}}{e-1} \left[ \Tr_N \left( -2e \hat\rho_N^2 + \left( \exp \frac{\hat\eta_k^\dagger \hat\eta_k}{2} \hat\rho_N \exp \frac{\hat\eta_k^\dagger \hat\eta_k}{2} \right)^2 + \left( \exp \frac{\hat\eta_k \hat\eta_k^\dagger}{2} \hat\rho_N \exp \frac{\hat\eta_k \hat\eta_k^\dagger}{2} \right)^2 \right) \right]^{1/2}.
\end{aligned}
\end{align}
\cref{eq:sm:rdm_exponential} establishes the exponential form of $\hat\rho_N$ while $\exp\small(\hat\eta_k^\dagger \hat\eta_k^\nodagger / 2\small)$ and $\exp \small(\hat\eta_k^\nodagger \hat\eta_k^\dagger / 2\small)$ are trivially exponential in Majorana operators $\hat a_j$.
The same can be shown for any product of such operators~\cite{Fagotti2010}
\begin{align}
    \exp \left( \frac{1}{4} \sum_{j,l = 0}^{2N-1} W_{j,l} \hat a_j \hat a_l \right) \exp \left( \frac{1}{4} \sum_{j,l = 0}^{2N-1} W_{j,l}^\prime \hat a_j \hat a_l \right) = \exp \left( \frac{1}{4} \sum_{j,l = 0}^{2N-1} [\ln (\exp(W) \exp(W^\prime))]_{j,l} \hat a_j \hat a_l \right).
\end{align}
The shape of \cref{eq:sm:coherence_trce_quead_exponentials}, which consequently contains only traces of squared exponential operators, is desirable because the identity~\cite{Vidal2003, Fagotti2010}
\begin{align} \label{eq:sm:trace_identity_one}
    \Tr \left( \hat\rho^2[\Gamma] \right) = \sqrt{\det \frac{1 + \Gamma^2}{2}}
\end{align}
can be exploited, where $\hat\rho[\Gamma]$ denotes a Hermitian, trace\hyp normalized and positive semi\hyp definite exponential of Majorana operators with correlation matrix $\Gamma$.
By employing Wick's theorem and $\exp\small(\hat\eta_k^\dagger \hat\eta_k^\nodagger / 2\small) = 1 + \hat\eta_k^\dagger \hat\eta_k^\nodagger (\sqrt{e} - 1)$, correlation matrices of the individual operators in \cref{eq:sm:coherence_trce_quead_exponentials} are fully determined by the correlation matrix of $\hat\rho_N$, \cref{eq:sm:rdm_correlation}.
Trace normalization, for example $\Tr_N \small(\exp \small( \hat\eta_k^\dagger \hat\eta_k^\nodagger \small) \hat\rho_N\small)$, is obtained from the generalization~\cite{Fagotti2010} of \cref{eq:sm:trace_identity_one}
\begin{align} \label{eq:sm:trace_identity_two}
    \Tr \left( \hat\rho[\Gamma] \hat\rho[\Gamma^\prime] \right) = \sqrt{\det \frac{1 + \Gamma \Gamma^\prime}{2}},
\end{align}
with both $\hat\rho[\Gamma]$ and $\hat\rho[\Gamma^\prime]$ being positive semi\hyp definite.
\newline \indent
We conclude this section with remarks on the choice of measured mode $k \in \{ 0, ..., N-1 \}$ and, in the next paragraph, on the quenched\hyp to parameter $\xi$.
When justifying our choice of \gls{CSO} measure, we have already mentioned it being independent on the orthonormal basis chosen to represent the density matrix within the subspaces corresponding to distinct measurement results.
This implies that, for a pure state $\hat\rho_N (0)$,
\begin{align}
    \coherence{\hat\rho_N (0)}{k} = 2 \sqrt{ \Tr_N (\hat\rho_N(0) \hat\Pi_k) \Tr_N (\hat\rho_N(0) (1 - \hat\Pi_k)) },
\end{align}
which takes its maximal value of $1$ if, and only if, $\Tr_N (\hat\rho_N \hat\Pi_k) = \Tr_N (\hat\rho_N (1 - \hat\Pi_k)) = 1/2$.
It is desirable to maximize this quantity because the observable difference to projective measurements becomes optimized for small inter\hyp measurement timescales.
\newline \indent
Local thermalization implies
\begin{enumerate*}[label=(\roman*)]
    \item $\lim_{t \to \infty} \coherence{\hat\rho_N (t)}{k}$ exists, and
    \item $\coherence{\hat\rho_N (t)}{k} - \lim_{t \to \infty} \coherence{\hat\rho_N (t)}{k} \propto t^{-3/2}$ as $t \to \infty$.
\end{enumerate*}
However, $\lim_{t \to \infty} \coherence{\hat\rho_N (t)}{k} = 0$ is not guaranteed.
We observe that decoherence resolves the measurement only for quenches not beyond the quantum critical point, while finite \gls{CSO} is approached for quenches into the ordered phase,
\begin{align}
    \lim_{t \to \infty} \coherence{\hat\rho_N (t)}{k}
    \begin{cases}
        = 0 ~ &\mathrm{if} ~ \xi \geq 1, \\
        > 0 ~ &\mathrm{if} ~ \xi < 1.
    \end{cases}
\end{align}
The limiting behavior is visualized in the main text and in \cref{sm:fig:initial_and_limiting_coherence}.
For quenches within the disordered phase, $\xi \geq 1$, the greatest initial \gls{CSO}, $\coherence{\hat\rho_N (0)}{k}$, is reached for $\xi = 1$.
\cref{sm:fig:initial_and_limiting_coherence} manifests this observation.
\section{A theory of decoherence\hyp based measurements} \label{sm:ch:decoherence_based_measurements}
Decoherence has been widely discussed in the context of the measurement problem, in particular in the context of describing measurements as continuous processes rather than instantaneous projections. 
If time evolution is purely described in terms of Schrödinger's equation, however, the measurement process has to be further specified.
While the measurement of states with vanishing \gls{CSO} is still equivalent to revealing the outcome of a classical probability experiment, the meaning of measuring coherent states is not clear a priori and the theory has to be extended to account for it.
First, we address a special class of states with maximal \gls{CSO}: equal superpositions of measurement outcomes, i.e. $(\ket{0} + \ket{1}) / \sqrt{2}$, where states $\ket{0}$ and $\ket{1}$ correspond to eigenstates of the measured observable with eigenvalues $0$ and $1$, respectively.
Clearly, an observer concludes either $0$ or $1$ to be an observation's result and also Born's rule for the corresponding probabilities can be taken for granted since decoherence relies thereon itself.
However, in the absence of wave\hyp function collapses, the post\hyp measurement state should still be $(\ket{0} + \ket{1}) / \sqrt{2}$ because any other such state would at least effectively imply some kind of time evolution beyond Schrödinger's equation, thus spoiling the desired unified formulation.
For a state with arbitrary \gls{CSO}, we exploit the convexity of the set of density operators and make the assumption that a measurement collapses all the fully incoherent information of the system's state.
We rewrite $\hat\rho_N (t) = \mathbb{1}/ 2^N + \mathbf{b} (t) \cdot \mathbf{\hat\Sigma} / 2$ where the components of $\mathbf{\hat\Sigma}$ are normalized orthogonal generators of $SU(N)$ such that $\Tr \hat\Sigma_j \hat\Sigma_l = 2 \delta_{jl}$.
It is well known that the set of density operators is defined by the Bloch\hyp space $\mathcal{B}(\mathbb{R}^{4^N - 1}) \equiv \{ \mathbf{b} = r\mathbf{e}_b \in \mathbb{R}^{4^N - 1} ~ | ~ 0 \leq r \leq 1 / (2^{N - 1} |m(\mathbf{e}_b \cdot \mathbf{\hat\Sigma})|)\}$ where $\mathbf{e}_b$ is a unit vector with respect to $l_2$ norm $\norm{\cdot}$ and $m(\mathbf{e}_b \cdot \mathbf{\hat\Sigma}) \equiv \min \sigma(\mathbf{e}_b \cdot \mathbf{\hat\Sigma})$ is the minimal value in the spectrum of $\mathbf{e}_b \cdot \mathbf{\hat\Sigma}$ ~\cite{Kimura2005}.
We choose to represent the \gls{RDM} as the convex combination with the largest possible contribution from the maximally mixed state
\begin{align} \label{eq:sm:convex_decomposition}
    \hat\rho_N (t) = 2^{N - 1} |m(\mathbf{b}(t) \cdot \mathbf{\hat\Sigma})| \norm{\mathbf{b}(t)} \hat\rho_N^B (t) + \left( 1 - 2^{N - 1} |m(\mathbf{b}(t) \cdot \mathbf{\hat\Sigma})| \norm{\mathbf{b}(t)} \right) \frac{\mathbb{1}}{2^N},
\end{align}
where $\hat\rho_N^B (t) \equiv (\mathbb{1} + \mathbf{e}_b (t) \cdot \mathbf{\hat\Sigma} / | m(\mathbf{e}_b (t) \cdot \mathbf{\hat\Sigma}) |) /2^N$ is the density operator on the boundary of the set of density operators in direction $\mathbf{b} (t)$.
The state $\hat\rho_N^B (t)$ is not necessarily pure itself but when decomposing it as a convex combination of the states that we have provided post\hyp measurement states for, the maximally mixed state cannot contribute non\hyp trivially.
We thus assume that a measurement leaves states on the boundary of the set of density operators invariant and measuring, for example, $\hat\Pi_k = 1$ at time $t_0$ causes only a classical collapse described by a quantum channel
\begin{align} \label{eq:sm:decoherence_post_measurement_state}
    \hat\rho_N (t_0) \to 2^{N - 1} |m(\mathbf{b}(t_0) \cdot \mathbf{\hat\Sigma})| \norm{\mathbf{b}(t_0)} \hat\rho_N^B (t_0) + \left(1 - 2^{N - 1} |m(\mathbf{b}(t_0) \cdot \mathbf{\hat\Sigma})| \norm{\mathbf{b}(t_0)} \right) \frac{\hat\Pi_k}{2^{N - 1}}.
\end{align}
Evidently, a similar mechanism could have been justified when decomposing $\hat\rho_N(t)$ into a boundary state and any state that exhibits vanishing \gls{CSO} with respect to the conducted measurement.
However, due to the fact that the maximally mixed state is the \textit{unique} description that exhibits no \gls{CSO} for any measurement, we argue that \cref{eq:sm:decoherence_post_measurement_state} describes the natural choice of theory.

It is also fruitful to adapt the view on decoherence as a system being continuously monitored by its environment~\cite{Schlosshauer2019} and understand \cref{eq:sm:convex_decomposition} as the separation of state information into fully captured by the environment ($\sim \mathbb{1}$) and inherent to the system ($\sim \hat\rho_N^B(t)$).
The partial classical collapse on the \gls{RDM}, \cref{eq:sm:decoherence_post_measurement_state}, corresponds to a non\hyp unitary operation on the pure \textit{global} state, which has to be specified in order to conclude the exact post\hyp measurement dynamics.
It is precisely this operation, however, that hints to the problems that decoherence cannot cure and leads to a wide field of possible theories.
We do not intend to promote any particular extension of decoherence\hyp based theories, but establish an upper bound that all of them should have in common.
In that order, it is important to note that it would be inconsistent to consider measurements as non\hyp fundamental while permitting their timing to affect a system's dynamics.
We thus conclude that the dynamics of the state information inherent to the open subsystem is described by $\hat\rho_N^B (t)$ throughout and irrespective of any interim measurements.
This argument implies that $2^{N - 1} |m(\mathbf{b}(t) \cdot \mathbf{\hat\Sigma})| \norm{\mathbf{b}(t)} \Tr \small( \hat\Pi_k \hat\rho_N^B (t) \small)$ is a lower bound for measuring $\hat\Pi_k = 1$ at any time $t$ and similarly for $\hat\Pi_k = 0$ independently of the measurement\hyp history.
We denote probabilities stemming from a decoherence point of view by $P_\mathrm{dec}$ and due to $|m(\mathbf{b} \cdot \mathbf{\hat\Sigma})| > 2^{1/2 - N} ~ \forall ~ \mathbf{b} \in \mathcal{B}(\mathbb{R}^{4^N - 1})$~\cite{Kimura2005} and $\norm{\mathbf{b} (t)} = \sqrt{2 \Tr_N \hat\rho_N^2 (t) - 2^{1-N}}$, we conclude
\begin{align} \label{eq:sm:decoherence_probability_bounds}
\begin{aligned}
    P_\mathrm{dec} (\hat\Pi_k (t) = 1) &> \max \left( 0, \frac{1}{2} \sqrt{\Tr_N \hat\rho_N^2 (t) - 2^{-N}} + \Tr_N (\hat\Pi_k \hat\rho_N (t)) - \frac{1}{2} \right), \\
    P_\mathrm{dec} (\hat\Pi_k (t) = 0) &> \max \left( 0, \frac{1}{2} \sqrt{\Tr_N \hat\rho_N^2 (t) - 2^{-N}} - \Tr_N (\hat\Pi_k \hat\rho_N (t)) + \frac{1}{2} \right).
\end{aligned}
\end{align}
\section{Practical measurements}
Due to its non\hyp local character, measuring the projector $\hat\Pi_k$ onto the subspace with occupied $k$\hyp mode in the open chain might be experimentally challenging.
Thus, here we discuss practical schemes that allow for the realization of the proposed measurement protocol.
\newline \indent
Arguably the most straightforward approach consists in measuring the occupations $n_j \in \{ 0, 1 \}$ of fermionic sites $j = 0, \dots, N-1$.
This corresponds to a measurement in the basis projected onto by $\hat\Pi_{\{ n_i \}} \equiv \prod_{i=0}^{N-1} (\hat c_i^\dagger \hat c_i^\nodagger)^{n_i} (\hat c_i^\nodagger \hat c_i^\dagger)^{(1-n_i)}$ with $\{ n_i \} \in \{ 0, 1 \}^N$.
Depending on the protocol's mode $k = 0, \dots, N - 1$,
\begin{align}
    \Tr_N \left( \hat\Pi_k \hat\Pi_{\{ n_i \}} \right) = \frac{1}{2} \left( 1 + \sum_{i=0}^{N-1} (-1)^{n_i} (h_{k,i}^2 - g_{k,i}^2) \right) \in [0, 1],
\end{align}
quantifies the overlap of the one\hyp dimensional subspace projected onto by $\hat\Pi_{\{ n_i \}}$ and the sought $2^{N-1}$\hyp dimensional subspace defined by $\hat\Pi_k = 1$, where $g_{k,i}$ and $h_{k,i}$ are defined in \cref{eq:sm:tfim_obc_mode_gh}.
Under the assumption of projective measurements and upon measuring $\hat\Pi_{\{ n_i^0 \}} = 1$ at time $t_0$, the probability to measure $\hat\Pi_k = 1$ instantaneously after the first measurement is $\Tr_N (\hat\Pi_k \hat\Pi_{\{ n_i \}})$ because $\rank \hat\Pi_{\{ n_i \}} = 1$.
For a second measurement at $t_1 = t_0 + t$, the probability to measure $\hat\Pi_k = 1$ reads
\begin{align} \label{eq:sm:practical_measurements_local_z_projective_prob}
\begin{aligned}
   P_\mathrm{proj}^\prime (\Pi_k (t_1) = 1) &= \Tr \left( \exp(-i\hat Ht) \left[ \hat\Pi_{\{ n_i^0 \}} \otimes \hat\rho_{L \backslash N} (t_0) \right] \exp(i\hat Ht) \hat\Pi_k \right) \\
   &\geq \Tr_N (\hat\Pi_k \hat\Pi_{\{ n_i^0 \}}) P_\mathrm{min} - \sqrt{(1 - \Tr_N (\hat\Pi_k \hat\Pi_{\{ n_i^0 \}})) \Tr_N (\hat\Pi_k \hat\Pi_{\{ n_i^0 \}})},
\end{aligned}
\end{align}
where $t \leq t_\mathrm{bound}$ has to hold for above inequality and the prime on $P_\mathrm{proj}^\prime$ indicates the measurement in the experimentally accessible basis.
The corresponding upper bounds for a decoherence\hyp based perspective derived from \cref{eq:sm:decoherence_probability_bounds} are unchanged and, consequently, the two perspectives contradict each other as long as $1 - \Tr_N (\hat\Pi_k \hat\Pi_{\{ n_i \}})$ is sufficiently small and $P_\mathrm{min}$ sufficiently large.
In order to obtain an experimental estimation of $P_\mathrm{proj}^\prime (\Pi_k (t_1) = 1)$, we follow a quantum\hyp state tomographic approach.
Rewriting
\begin{align}
    \hat\rho_N(t_1) = \sum_{\{ n_i \}} \Tr_N \left( \hat\Pi_{\{ n_i \}} \hat\rho_N(t_1) \right) \hat\Pi_{\{ n_i \}}
\end{align}
and performing $R$ measurements on independent systems at $t_1$ following the initial measurement of $\Pi_{\{ n_i^0 \}} = 1$ at time $t_0$, the respective parameters can be estimated as
\begin{align}
    \Tr_N \left( \hat\Pi_{\{ n_i \}} \hat\rho_N(t_1) \right) \approx \frac{\# \text{measurements with} ~ \Pi_{\{ n_i \}} = 1}{R}.
\end{align}
As usual, the standard deviation of this estimation scales $\propto R^{-1/2}$.
An estimation of the desired probability is obtained by using the above expression for the exact result
\begin{align} \label{eq:sm:practical_measurements_prob_pik1}
    \Tr_N (\hat\rho_N(t_1) \hat\Pi_k) = \sum_{\{ n_i \}} \Tr_N \left( \hat\Pi_{\{ n_i \}} \hat\rho_N(t_1) \right) \Tr_N \left( \hat\Pi_{\{ n_i \}} \hat\Pi_k \right).
\end{align}
As an example, consider the proposed protocol for $N=8$.
We have $\Tr_N ( (\mathbb{1} - \hat\Pi_1) \hat\Pi_{\{ n_i^0 \}} ) > 0.87$ if sites $i = 0, ..., 7$ are unoccupied on first measurement except for $n_3^0 = n_4^0 = 0$.
According to \cref{eq:sm:practical_measurements_local_z_projective_prob}, this implies a bound of $P_\mathrm{proj}^\prime(\Pi_1(t_1) = 0) > 0.52$, which is in clear contradiction to decoherence\hyp based measurements, as we show in Fig. 2 in the main text.
Again for $N = 8$ but with different choices of $k$ and $\{n_i^0\}$, probabilities of $P_\mathrm{proj}^\prime (\Pi_k(t_1) = \pm 1) \approx 0.9$ are easily obtained, which comes at the cost of smaller initial \glspl{CSO}, however, and correspondingly greater upper bounds on $P_\mathrm{dec}$.
\newline \indent
In light of the \gls{JW} transformation \cref{eq:sm:jw_trafo}, measuring the occupation of fermionic sites relates to measuring spin components along the $z$\hyp axis.
The local character of the measurement, however, is preserved for projections along any axes.
We may define unit vectors $\mathbf{d}_i \equiv (\sin \theta_i \cos \varphi_i, \sin \theta_i \sin \varphi_i, \cos \theta_i)$ that parametrize the axes for spin $i=0, \dots, N-1$.
Correspondingly, the projector onto the one\hyp dimensional subspace for which spin $i$ is in the eigenstate of $\mathbf{d}_i \cdot \mathbf{\hat\sigma}_i$ with eigenvalue $n_i \in \{ \pm 1 \}$ is
\begin{align} \label{eq:sm:practical_measurement_flexible_basis}
    \hat\Pi_{\{ n_i \}} (\{ \theta_i, \varphi_i \}) \equiv \prod_{i = 0}^{N-1} \hat\Pi_{\mathbf{d}_i, n_i},
\end{align}
where we have defined the projectors onto the $n_i = 1$ eigenstate
\begin{align}
    \hat\Pi_{\mathbf{d}_i, 1} \equiv \cos \theta_i \hat c_i^\dagger \hat c_i^\nodagger + \sin^2 \frac{\theta_i}{2} + \frac{\sin \theta_i}{2} \exp (i\pi \sum_{j<i} \hat c_j^\dagger \hat c_j^\nodagger) \left( e^{i \varphi_i} \hat c_i + e^{-i \varphi_i} \hat c_i^\dagger \right),
\end{align}
and $\hat\Pi_{\mathbf{d}_i, -1} \equiv \mathbb{1} - \hat\Pi_{\hat{\mathbf{d}}_i, 1}$.
The additional degrees of freedom $\{ \theta_i, \varphi_i \}$ allow for optimizing the measurement in the sense of extremizing the overlap of the subspaces $\hat\Pi_{\{ n_i \}} (\{ \theta_i, \varphi_i \}) = \pm 1$ with the subspaces of well\hyp defined occupation of mode $k$.
In light of \cref{eq:sm:practical_measurements_local_z_projective_prob}, one may formulate this $k$\hyp dependent optimization problem as $\{ \theta_i^\mathrm{opt}, \varphi_i^\mathrm{opt} \} = \mathrm{argmin}_{\{ \theta_i, \varphi_i \}} L (\{ \theta_i, \varphi_i \})$ with
\begin{align} \label{eq:sm:practical_measurements_optimization_option_2}
    L (\{ \theta_i, \varphi_i \}) \equiv \sum_{\{ n_i \}} \Tr_N(\hat\Pi_k \hat\Pi_{\{ n_i \}} (\{ \theta_i, \varphi_i \})) \left(1 - \Tr_N(\hat\Pi_k \hat\Pi_{\{ n_i \}} (\{ \theta_i, \varphi_i \}))\right) \in [0, 2^{N-2}].
\end{align}
\newline \indent
Repeated measurements amplify the differences between projective and collapse\hyp free theories.
So far, we have only considered practical realizations of our protocol for one confirming measurement following the initial one.
Now, we discuss how to extend the described procedure of measuring in an approximate basis to protocols of multiple confirming observations.
With an increasing amount of confirmations, it becomes more important that the basis states projected onto by $\hat\Pi_{\{ n_i \}} (\{ \theta_i, \varphi_i \})$ have large overlaps with either the $\Pi_k = 1$ or $\Pi_k = 0$ subspaces.
This is due to the fact that collapses into states with large components from both $\Pi_k = 0, 1$ subspaces weakens the ability to make conclusions about subsequent measurements of $\hat\Pi_k$.
Let us assume that we have optimized the basis such that $\Tr_N(\hat\Pi_k \hat\Pi_{\{ n_i \}} (\{ \theta_i, \varphi_i \})) \in [0,1] \backslash (\epsilon, 1-\epsilon) ~ \forall \{n_i\} \in \{0,1\}^N$ and some $\epsilon \in [0, 1/2)$.
If insisting on repeated measurements, the optimization problem would thus be formulated with $L (\{ \theta_i, \varphi_i \}) = \epsilon$ instead of \cref{eq:sm:practical_measurements_optimization_option_2}.
We then define the orthorgonal $2^{N-1}$\hyp dimensional subspaces $S_{N,k}^{+ (-)}$ of $\mathcal{H}_N$ as the direct sum of one\hyp dimensional subspaces projected onto by $\hat\Pi_{\{ n_i \}} (\{ \theta_i, \varphi_i \})$ with $\Tr_N (\hat\Pi_k \hat\Pi_{\{ n_i \}} (\{ \theta_i, \varphi_i \})) \in [1-\epsilon, 1] ~ ([0, \epsilon])$.
That means for $\hat\Pi_{\{ n_i \}} (\{ \theta_i, \varphi_i \}) : \mathcal{H}_N \to S_{N,k}^+$ that \cref{eq:sm:practical_measurements_local_z_projective_prob} can be further bounded as $P_\mathrm{proj}^\prime (\Pi_k (t_1) = 1) \geq (1 - \epsilon) P_\mathrm{min} - \sqrt{\epsilon (1 - \epsilon)}$.
We now have to derive a lower bound for the probability $P_\mathrm{proj}^\prime (\Pi_k(t_n) = 1)$ that, given an initial measurement of the state projected onto by $\hat\Pi_{\{ n_i \}} (\{ \theta_i, \varphi_i \}) : \mathcal{H}_N \to S_{N,k}^+$, one would measure $\Pi_k = 1$ on the $n^\mathrm{th}$ confirming measurement, assuming wave function collapses (similarly we could consider $\hat\Pi_{\{ n_i \}} (\{ \theta_i, \varphi_i \}) : \mathcal{H}_N \to S_{N,k}^-$ and $\Pi_k = 0$).
The intermediate measurements, however, do not precisely project the state into $\Pi_k = \pm 1$ subspaces but into $S_{N,k}^\pm$ instead.
Noting that we assume a large overlap between these as quantified by small $\epsilon$ and that \cref{eq:sm:practical_measurements_local_z_projective_prob} consequently implies large overlap between the $\Pi_k = 1$ subspace and a state that has been in $S_{N,k}^+$ time $t$ ago, we bound $P_\mathrm{proj}^\prime (\Pi_k(t_n) = 1)$ from below via the probability that for every measurement $1,\dots, n-1$ a state in $S_{N,k}^+$ is observed.
Thus, in the first place, we need a lower bound for the probability that a measurement of a state being in $S_{N,k}^+$ is confirmed in the subsequent observation.
Using \cref{eq:sm:practical_measurements_local_z_projective_prob} and $\hat\rho_N(t_j) = \exp (-i\hat H t) [ \hat\Pi_{\{ n_i^{j-1} \}} \otimes \hat\rho_{L \backslash N} (t_{j-1}) ] (\{ \theta_i, \varphi_i \}) \exp (i\hat H t)$ with $\hat\Pi_{\{ n_i^{j-1} \}} (\{ \theta_i, \varphi_i \}) : \mathcal{H}_N \to S_{N,k}^+$, one can derive
\begin{align} \label{eq:sm:practical_measurements_remain_splus}
    \Tr_N (\hat\rho_N(t_j) \hat\Pi_{S_{N,k}^+}) \geq \left[ (1 - \epsilon) P-\mathrm{min} - \sqrt{\epsilon (1 - \epsilon)} \right] \min_{\ket{v} : \hat\Pi_k \ket{v} = \ket{v}} \braket{v|\hat\Pi_{S_{N,k}^+}|v} - 2 \sqrt{1 - (1-\epsilon) P_\mathrm{min} + \sqrt{\epsilon (1-\epsilon)}},
\end{align}
defining $\hat \Pi_{S_{N,k}^+} : \mathcal{H}_N \to S_{N,k}^+$ as the projector onto $S_{N,k}^+$ of rank $2^{N-1}$ and requiring $t \leq t_\mathrm{max}$.
Lastly, the above summand involving the minimum is to be bounded from below, which we do dependent on $N$ since $\dim \mathcal{H}_N = 2^N$.
By assumption, for $N=1$ we have $\min_{\hat\Pi_k \ket{v} = \ket{v}} \braket{v|\hat\Pi_{S_{1,k}^+}|v} \geq 1-\epsilon$ and using Cauchy\hyp Schwarz inequality one may derive the recursion
\begin{align}
    \left. \min_{\ket{v} : \hat\Pi_k \ket{v} = \ket{v}} \braket{v|\hat\Pi_{S^+}|v} \right|_{\dim \mathcal{H} = 2^N} \geq \left. \min_{\ket{v} : \hat\Pi_k \ket{v} = \ket{v}} \braket{v|\hat\Pi_{S^+}|v} \right|_{\dim \mathcal{H} = 2^{N-1}} + \left( \left. \min_{\ket{v} : \hat\Pi_k \ket{v} = \ket{v}} \braket{v|\hat\Pi_{S^+}|v} \right|_{\dim \mathcal{H} = 2^{N-1}} \right)^2 - 1.
\end{align}
As an example, if we reach $\epsilon = 10^{-4}$ and $P_\mathrm{min} = 0.99$ for $N = 8$ spins in the subsystem, then \cref{eq:sm:practical_measurements_remain_splus} would read $\Tr_N (\hat\rho_N(t_j) \hat\Pi_{S_{N,k}^+}) > 0.62$.
Accordingly, $P_\mathrm{proj}^\prime (\Pi_k (t_2) = 1) > 0.61, ~ P_\mathrm{proj}^\prime (\Pi_k (t_3) = 1) > 0.37$ and $P_\mathrm{proj}^\prime (\Pi_k (t_n) = 1 | n = 1,2,3) > 0.22$, again in clear contradiction to a decoherence\hyp based theory, see main text.
\newline \indent
The previous example made clear that, for increasing $N$, increasingly small values of $\epsilon$ are necessary in order to obtain non\hyp trivial bounds for $P_\mathrm{proj}^\prime$.
By measuring $\hat\Pi_{\{ n_i \}} (\{ \theta_i, \varphi_i \})$ and thus purely relying on local information, such good agreement among the $\Pi_k = \pm 1$ subspaces and $S_{N,k}^\pm$ may, however, be out of reach.
If this is the case, we may define orthorgonal subspaces $\tilde{S}_{N,k}^{+ (-)}$ that are similarly defined as the direct sums of one\hyp dimensional subspaces projected onto by $\hat\Pi_{\{ n_i \}} (\{ \theta_i, \varphi_i \})$ with $\Tr_N (\hat\Pi_k \hat\Pi_{\{ n_i \}} (\{ \theta_i, \varphi_i \})) \in [1-\epsilon, 1] ~ ([0, \epsilon])$ but now $\mathcal{H}_N \neq \tilde{S}_{N,k}^+ \oplus \tilde{S}_{N,k}^-$.
In other words, we optimize the measured basis \cref{eq:sm:practical_measurement_flexible_basis} such that, for $\epsilon$ fixed, $\dim \tilde{S}_{N,k}^\pm$ is maximized but they are no longer required to be of dimension $2^{N-1}$.
The previously discussed bound \cref{eq:sm:practical_measurements_remain_splus}, however, becomes invalid for this setting as a state that has been observed to be in $\tilde{S}_{N,k}^+$ time $t$ ago is still likely to be found in the $\Pi_k = 1$ subspace but not necessarily in $\tilde{S}_{N,k}^+$ anymore.
However, it is in fact still unlikely to be found in $\tilde{S}_{N,k}^-$.
Thus, for the $n^\mathrm{th}$ measurement we can only make conclusions if at all interim measurements $1, \dots, n-1$ states in either $\tilde{S}_{N,k}^+$ or $\tilde{S}_{N,k}^-$ were observed.
This reduces the acceptance probability for the protocol but $\Tr_N (\hat\rho_N(t_j) \hat\Pi_{\tilde{S}_{N,k}^+})$ can again be bounded from below similarly to \cref{eq:sm:practical_measurements_remain_splus} where the bound for the summand involving the minimum can be adapted according to the dimension of $\tilde{S}_{N,k}^\pm$.
\end{document}